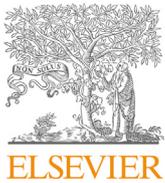
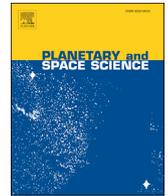

# Complementary astrometry of *Cassini* Imaging Science Subsystem images of phoebe

Q.F. Zhang [a,b,*], W.H. Qin [a], Y.L. Ma [c], V. Lainey [d], N.J. Cooper [e], N. Rambaux [d], Y. Li [a,b], W.H. Zhu [a,b]

[a] *Department of Computer Science, Jinan University, Guangzhou, 510632, P. R. China*
[b] *Sino-French Joint Laboratory for Astrometry, Dynamics and Space Science, Jinan University, Guangzhou, 510632, P. R. China*
[c] *School of Data Science and Engineering, East China Normal University, Shanghai, 200062, P. R. China*
[d] *IMCCE, CNRS, Observatoire de Paris, PSL Université, Sorbonne Université, Université de Lille 1, UMR 8028 du CNRS, 77 Denfert-Rochereau 75014 Paris, France*
[e] *Astronomy Unit, Department of Physics > Astronomy, Queen Mary University of London, Mile End Road, London, E1 4NS, UK*

A B S T R A C T

Phoebe is the only major satellite of Saturn with a retrograde orbit. The *Cassini* Imaging Science Subsystem (ISS) took a lot of Phoebe images between 2004 and 2017, but only a selection of them has been reduced. In this paper, we reduced the remaining ISS images of Phoebe. In the reduction, the Gaia EDR3 catalogue was used to provide the reference stars' positions, and the modified moment was used to measure the centre of image stars and Phoebe. Finally, a total of 834 ISS images of Phoebe have been reduced successfully. Compared with the JPL ephemeris SAT375, Phoebe's positions are consistent. The average residuals in the right ascension and declination are 0.08″ and −0.05″, and the standard deviations of the residuals are about 0.2″. In terms of residuals in linear units, the means in the right ascension and declination are about 5 km and −2 km, respectively; The standard deviations are about 11 km. Compared with the JPL ephemeris SAT427 and IMCCE ephemeris PH20, our measurements show a strong bias and a large dispersion.

## 1. Introduction

An optical Imaging Science Subsystem (ISS) (Porco et al., 2004) mounted on the *Cassini* orbiter has taken more than 440,000 images. These images have become an important resource for the astrometry of natural satellites. For example, Cooper et al. (2006) reduced ISS images of inner Jovian satellites; Cooper et al. (2014) performed mutual-event astrometry of ISS images of the mid-sized Saturnian satellites. A selection of ISS images of some main icy Saturnian satellites has been reduced by Tajeddine et al. (2013, 2015) and Zhang et al. (2018a,b). In 2018, Cooper et al. (2018) released a software package to the community, *Caviar* (https://www.imcce.fr/recherche/equipes/pegase/caviar), which is dedicated to the astrometric reduction of *Cassini* ISS images. Recently, Lainey et al. (2020) combined the high-precision ISS images' astrometric results with *Cassini* radio science data to show that Titan is moving away from Saturn at a faster pace, implying that Saturn is one order of magnitude more tidally dissipative than previously thought. These researches demonstrated that high-precision astrometric data can be obtained from ISS images and play important roles in relevant fields.

Phoebe is the largest irregular Saturnian satellite. Many researchers are interested in its physical properties. Simonelli et al. (1999) measured its albedo, Castillo-Rogez et al. (2012) discussed its geophysical evolution, Fraser and Brown (2018) estimated its surface composition, Rambaux and Castillo-Rogez (2020) analysed its global shape. Astrometry is also important. Veiga et al. (2000) delivered their observations of Phoebe between 1995 and 1997. Peng et al. (2004, 2012, 2015) and Peng and Zhang (2006) developed a series of methods to measure the position of Phoebe from their observations. Qiao et al. (2006, 2011) also published their observations of Phoebe between 2003 and 2008. Tajeddine et al. (2015) reduced *Cassini* ISS images of Phoebe taken in June 2004. Gomes-Júnior et al. (2015) obtained more than 8000 astrometric positions of 18 irregular satellites of giant planets from 1992 to 2014, including 1787 observations of Phoebe. Gomes-Júnior et al. (2020) reported the first observations of stellar occultations by Phoebe between mid-2017 and mid-2019, and greatly improved the rotational period of Phoebe. At the same time, they gave six astrometric positions of Phoebe with very high precision of 1-mas level. All ground and space-based observations of Phoebe advanced the update of the ephemeris of Phoebe (Arlot et al., 2003; Jacobson, 2004; Shen et al., 2005, 2011; Emelyanov, 2010; Desmars et al., 2013). All these researches show the importance of the astrometry of Phoebe.

*Cassini* ISS provided a lot of observations of Phoebe between 2004 and 2017. As stated above, Tajeddine et al. (2015) reported the astrometry of ISS images of Phoebe taken between June 6–12, 2004. However, other observations of Phoebe have not been reduced. In this paper, we reduced the remaining ISS images of Phoebe between 2004 and 2017. In Section 2, the observations are introduced. In Section 3, the steps of reduction are detailed and the results are displayed. In Section 4, the results are analysed. In Section 5, the conclusions are drawn.

## 2. Observations

We gathered all ISS images of Phoebe taken between 2004 and 2017 from the Planetary Data System (https://pds-imaging.jpl.nasa.gov/). The total number of Phoebe images is 1880, including 1651 images taken by the ISS Narrow Angle Camera (NAC) and 229 images taken by the Wide Angle Camera (WAC). Through careful checking, we discarded all WAC images because either Phoebe's signal is very weak, or there are no reference stars in them. In all NAC images of Phoebe, firstly, those reduced by Tajeddine et al. (2015) have been excluded because they have been measured and the centring technique used in this paper is not suitable for them. Then, some images with poor quality due to the short exposure duration, noise corruption, scattered light and so on have been discarded. Finally, a total of 834 ISS NAC images of Phoebe were successfully reduced. The specifications of NAC (Porco et al., 2004) are listed in Table 1. All these images were taken between 2004 and 2007 and in 2015 (Table 2). These images' exposure duration ranges from 0.12 s to 26 s. Their solar phase angles vary from $2°$ to $140°$.

In every image measurement, Phoebe is unresolved and displayed as a point-like object. During the *Cassini* tour, Phoebe is resolved only in the NAC images taken between June 6–12, 2004 when *Cassini* was performing a Phoebe flyby. Fig. 1 gives an example image. In the image, Phoebe is only a point-source whose image size is several pixels. The highest resolution of Phoebe in all our reduced images is about 20km/pixel, and its apparent area is about $10 \times 10$ pixels. Obviously, a centring method is suitable to obtain Phoebe's centre instead of a limb-fitting method.

NAC has dual filters to support different exploration aims. Generally, the images taken with filters combination of clear filter 1 (CL1) and clear filter 2 (CL2) (Hereafter (CL1, CL2)) are used for astrometry because (CL1, CL2) has the best capability of detecting faint stars (Porco et al., 2004). In this paper, we carried out astrometry on all ISS images taken with different filter combinations. The influence of filter combinations on positional measurement is discussed in Section 4.

## 3. Data reduction

We used the dedicated package of the astrometry of ISS image, *Caviar*, to reduce all ISS images of Phoebe. The whole reduction procedure includes three main steps: pointing correction, target centring and target centre's conversion. The details are given below.

(1) In pointing correction, the nominated pointing of the ISS camera is corrected by matching image stars with catalogue stars. At first,

**Table 1**
The specifications of the ISS Narrow Angle Camera.

| Parameters | Values |
|---|---|
| Focal length | $2002.70 \pm 0.07$ mm |
| Pixel angular size | 5.9907 $\mu r$/pixel |
| FOV | 6.134 mrad |
| FWHM of PSF | 1.3 pixels |
| Filters | $12 \times 2$ filter wheels |
| Limiting magnitudes (in exposure time = 1s) | $M_v \sim 14$ |

**Table 2**
The distribution of all available observations of Phoebe.

| Year | 2004 | 2005 | 2006 | 2007 | 2015 | overall |
|---|---|---|---|---|---|---|
| Number of images | 190 | 216 | 124 | 5 | 299 | 834 |

we detect all possible star-like objects in an ISS image and compute their centres by the modified moment method. They are referred to as image stars. Secondly, all image stars' coordinates are corrected by the geometric distortion model given in Owen (2003). Thirdly, all possible stars in the field of view (FOV) of NAC are extracted from Gaia Early Data Release 3 (Gaia EDR3) (Gaia Collaboration, 2016, 2021) according to the nominated pointing of the camera, and their image coordinates are derived from their ICRS celestial coordinates in Gaia under the help of the NAIF (Navigation and Ancillary Information Facility) SPICE library (Acton (1996); Acton et al. (2018)). In the process, these celestial positions are corrected for stellar aberration and proper motions. Finally, the catalogue stars are matched with the image stars through their image coordinates to get reference stars, and the camera's pointing is corrected by the least-square method based on these reference stars' positions. It should be noted that the pointing correction is more accurate than before in *Caviar* because Gaia EDR3 and the modified moment method provide more accurate positions for catalogue stars and image stars, respectively.

(2) In target centring, the modified moment method is applied to obtain Phoebe's centre of light (COL). In all these available NAC images, Phoebe's resolution has a large range that varies from $\sim$ 20km/pixel to $\sim$ 450km/pixel. In a few images with the highest resolution, Phoebe's apparent size is only about $10 \times 10$ pixels. So centring method is suitable. Zhang et al. (2021) point out the modified moment is better than Gaussian fitting for the measurement of point-like objects due to the under-sampled feature of ISS NAC. Hence the modified moment method is used. In addition, the modified moment method is more robust than 2D Gaussian-fitting because the Gaussian fitting process will fail and can not obtain its centre when the target is very faint. Therefore, we added the modified moment method into *Caviar* to provide a more accurate centre for a point source. After that, we performed a phase correction (Lindegren (1977); Cooper et al. (2006)) on all Phoebe's COLs. The solar phase angles have a big range ($2°$–$140°$), so the phase correction is necessary. Finally, we obtain the image coordinates of centre of Phoebe with phase correction in each observation.

(3) In the target centroid conversion, the image coordinates of Phoebe's centre are converted to ICRS celestial coordinates centred at *Cassini*. The conversion includes one scale transformation and one inverse gnomonic projection. For further details, see also Cooper et al. (2006).

Eventually, through the careful reduction, a total of 834 Phoebe positions have been derived. All results are given as Table 3. Each row gives a Phoebe observation. The first column shows the reduced ISS image's ID. The second column is the middle time of the exposure when the image has been taken. Columns 3 $\sim$ 5 show the camera's pointing vector when taking the image. Columns $\alpha$ and $\delta$ are Phoebe's ICRS celestial coordinates centred at *Cassini*. Its corresponding image positions are given in columns Column and Line. It should be noted that each Phoebe position is corrected for solar phase effect. For convenience, the positions of Phoebe without phase correction are displayed in columns $\alpha_0$, $\delta_0$, $Column_0$ and $Line_0$. The numbers of reference stars are given in column $N_{ref}$. Finally, considering the filters' possible effect, we list the filter combination used by NAC for each image in column Filters, which will benefit users to evaluate these data. For the detailed specifications of all filter combinations, see also Porco et al. (2004). The full Table 3 is available in CDS (https://cdsarc.unistra.fr/viz-bin/cat/J/other/P+SS).

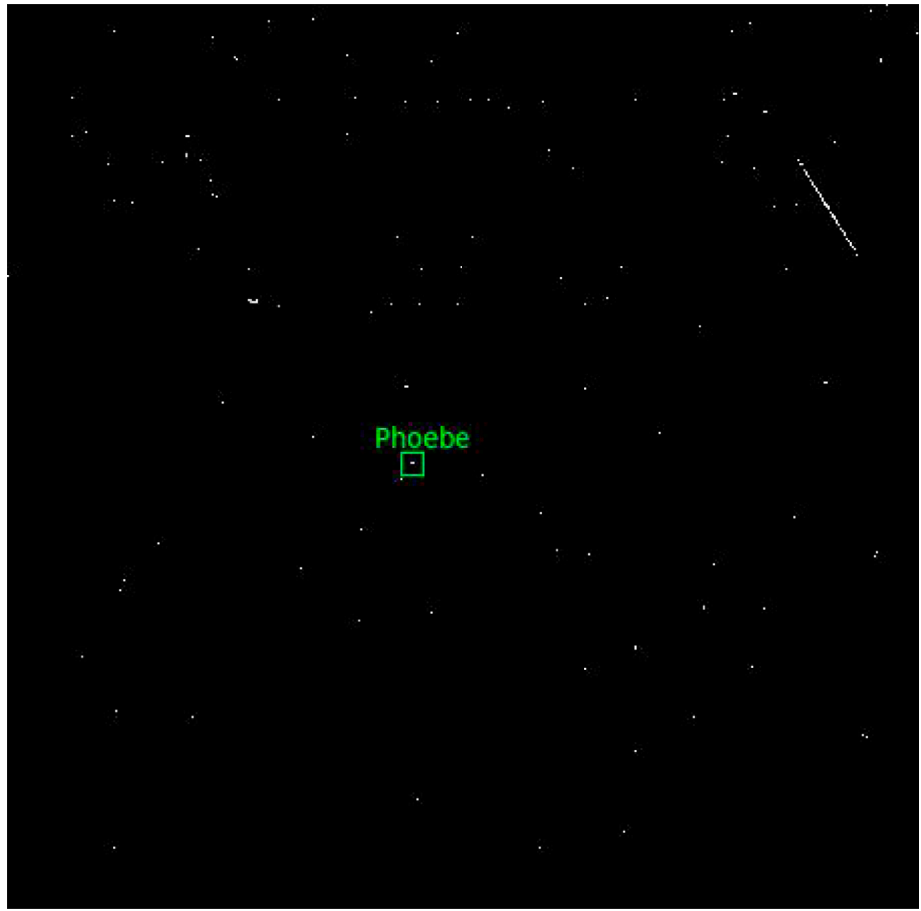

**Fig. 1.** NAC image N1454728139. Phoebe is a point source marked by a green square box. The image has been modified using a log transformation for the visibility of Phoebe. The exposure duration is 1*s*, and filter combination is (CL1, CL2). The resolution is about 450km/pixel.

**Table 3**
Sample of Phoebe's *Cassini* ISS observations. Column 1 is the *Cassini* ISS image ID. Column 2 is the date and exposure mid-time of the image (UTC). Columns $\alpha_c$, $\delta_c$, and Twist refer to the right ascension, declination, and twist angle of the camera's pointing vector in ICRS centred at *Cassini*, while $\alpha$ and $\delta$ are the right ascension and declination in ICRS centred at *Cassini* for Phoebe after phase correction. The columns of Column and Line are the observed positions with phase correction of Phoebe in the image. Columns $\alpha_0$ and $\delta_0$ are right ascension and declination in ICRS before phase correction. Columns $Column_0$ and $Line_0$ are the observed positions of Phoebe in image before phase correction. Column $N_{ref}$ is the number of reference stars. Column Filters is the filter combination with which the ISS NAC taken. The full table is available from the CDS. The origin of the column, line coordinate system is at the top left of the image, and line *y* increasing downwards and column *x* to the right. All the angle variables are given in degrees. Image size is 1024 × 1024 pixels or 512 × 512 pixels.

| Image_ID | Mid_Time (UTC) | $\alpha_c$ (deg) | $\delta_c$ (deg) | Twist (deg) | $\alpha$ (deg) | $\delta$ (deg) | Column (px) | Line (px) |
|---|---|---|---|---|---|---|---|---|
| N1455053281 | 2004/Feb/09 21:05:05.861 | 28.0091833 | 6.2006001 | 178.6552018 | 27.9652672 | 6.2536956 | 642.258 | 359.848 |
| N1460164774 | 2004/Apr/09 00:56:05.568 | 22.8622798 | 3.9976187 | 88.0980734 | 22.8640982 | 3.9704576 | 432.603 | 503.596 |
| N1509948065 | 2005/Nov/06 05:32:06.460 | 264.6732214 | −26.3847099 | 270.8454169 | 264.6706201 | −26.3874363 | 519.336 | 504.609 |
| N1515085437 | 2006/Jan/04 16:34:31.933 | 197.0137260 | −5.2499540 | 272.7240886 | 197.0135814 | −5.2524048 | 518.611 | 510.768 |
| N1799293550 | 2015/Jan/07 02:45:05.284 | 206.5882273 | −7.5070673 | 28.2289595 | 206.5093706 | −7.5229315 | 144.511 | 289.238 |

Sample of Phoebe's *Cassini* ISS observations.

| Image_ID | $\alpha_0$ (deg) | $\delta_0$ (deg) | $Column_0$ (px) | $Line_0$ (px) | $N_{ref}$ | Filters |
|---|---|---|---|---|---|---|
| N1455053281 | 27.9652267 | 6.2536796 | 642.374 | 359.897 | 10 | ('CL1','CL2') |
| N1460164774 | 22.8640091 | 3.9704217 | 432.490 | 503.851 | 17 | ('CL1','CL2') |
| N1509948065 | 264.6711168 | −26.3874305 | 519.339 | 505.905 | 41 | ('P60','GRN') |
| N1515085437 | 197.0139112 | −5.2525400 | 519.050 | 511.705 | 10 | ('CL1','IR1') |
| N1799293550 | 206.5093206 | −7.5229111 | 144.461 | 289.298 | 7 | ('CL1','GRN') |

## 4. Analysis of observations

*4.1. Comparison with three ephemerides*

To analyse these observations, we compared the reduced results with different ephemerides of Phoebe. The compared ephemerides include the current Institut de Mécanique Céleste et de Calcul des Ephémérides (IMCCE) ephemeris PH20 given by Desmars et al. (2013) (https://ftp.imcce.fr/pub/ephem/satel/phoebe/ph20.bsp), the current JPL ephemeris SAT427 (https://naif.jpl.nasa.gov/pub/naif/generic_kernels/spk/satellites/sat427.bsp) and the earlier JPL ephemeris SAT375 (https://naif.jpl.nasa.gov/pub/naif/generic_kernels/spk/satellites/a_old_versions/sat375.bsp).

At first, we computed the pixel coordinates of Phoebe in each image and the celestial coordinates of Phoebe in ICRS centred at *Cassini* from these ephemerides. Then we compared these calculated positions with the observed positions of Phoebe including phase correction to obtain the Observation-minus-Calculated (O–C) residuals. The (O–C)s in column and line relative to the three ephemerides are shown in Fig. 2. The (O–C)s in right ascension and declination are shown in Fig. 3 and Fig. 4. The former shows them in arc seconds, and the latter in km. Table 4 gives relevant statistics.

These figures and tables show that there exists a systematic difference among the three different ephemerides. Table 4 outlines the difference. The residuals relative to SAT375 are significantly smaller than those relative to PH20 and SAT427. For example, the mean residuals relative to SAT375 reach 5.3 and −2.2 km in right ascension and declination, respectively. The standard deviations are 11.3 and 10.9 km, respectively. However, relative to PH20 and SAT427, the corresponding values are several times greater. The best fitting of the measurements is SAT375 among the three ephemerides.

*4.2. Filters*

Of all 834 observations, 351 images were taken with CL1 and CL2, and the remaining images were taken with a few other filter combinations. According to whether the images of Phoebe were taken with (CL1, CL2), we divided the observation positions of Phoebe into two classes: (CL1, CL2) and non (CL1, CL2). The relevant statistics are displayed in Table 5. The table shows that the data from non (CL1, CL2) has a slightly greater standard deviation than those from (CL1, CL2). It indicates non (CL1, CL2) probably produce slightly bigger errors. It should be noted that the exposure duration of an image with non (CL1, CL2) was often longer than that with (CL1, CL2). It probably increased the positional error. But it is not entirely clear because the observation conditions varied during the observation period, and the positional uncertainty can be caused by many conditions, not only filter combinations. It only indicates that data from non (CL1, CL2) should be used carefully.

*4.3. Error sources*

For the error sources, we should note the following points.

(1) All three ephemerides give the centre of mass (COM) of Phoebe, but the measurements give the COL of Phoebe. When obtaining the residuals, we compare observed COL with computed COM. Generally, we assume that the COL, the centre of figure (COF) and COM are consistent; they are the same point. However, Phoebe's shape is irregular, its surface is heavily cratered and the albedo on its surface has a large variation (Porco et al., 2005). These features indicate that the COL deviates from its COF and COM. Hence, the difference between COM and COL produces larger residual errors.
(2) Phase correction models affect the accuracy of Phoebe's position. Fig. 5 shows the distribution of the residuals of Phoebe's positions in column and line over solar phase angle. From Fig. 5(a) and (b), it can be found that small solar phase angle (for example, < 20°)

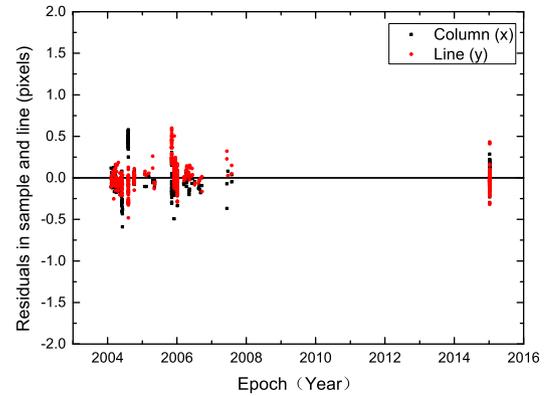

(a) Residuals relative to SAT375

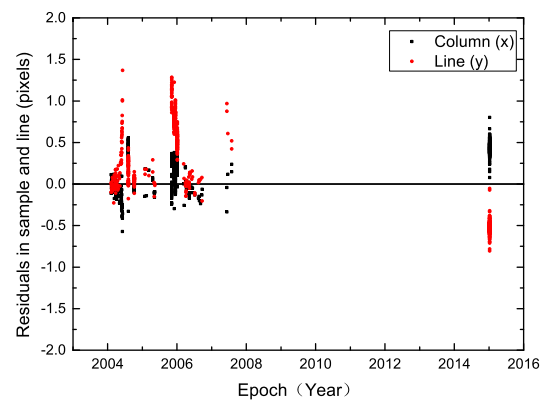

(b) Residuals relative to SAT427

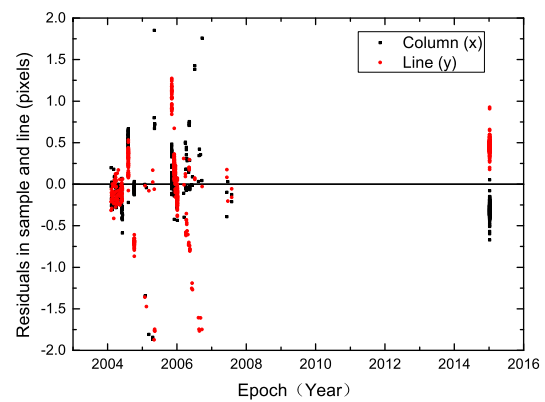

(c) Residuals relative to PH20

**Fig. 2.** The (O–C)s of positions of Phoebe in column and line relative to three different ephemerides: SAT375, SAT427 and PH20.

has a small residual. On the contrary, the residuals become great when it is big (for example, > 30°). The rule fails in Fig. 5(c), but it

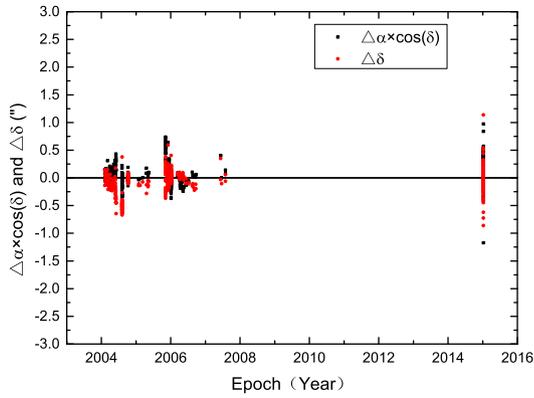

(a) Residuals relative to SAT375

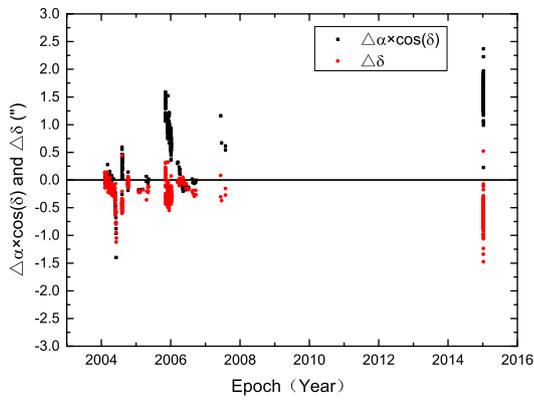

(b) Residuals relative to SAT427

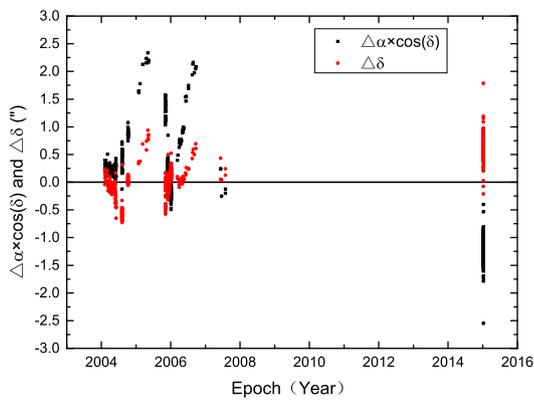

(c) Residuals relative to PH20

**Fig. 3.** The (O–C)s of positions of Phoebe in right ascension and declination relative to three different ephemerides: SAT375, SAT427 and PH20. Unit in arcsecond.

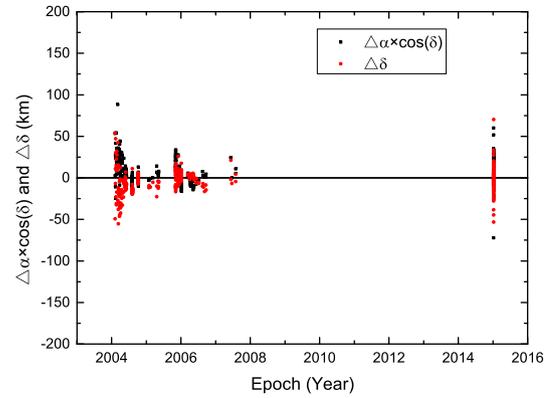

(a) Residuals relative to SAT375

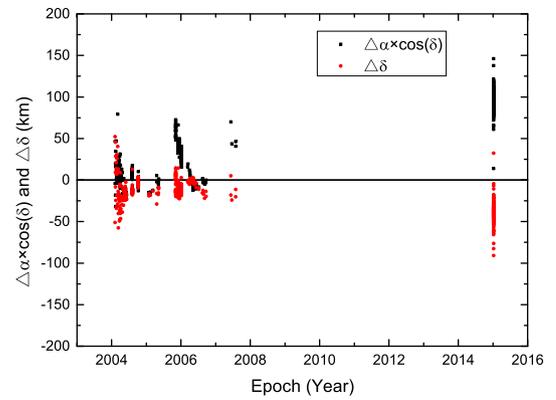

(b) Residuals relative to SAT427

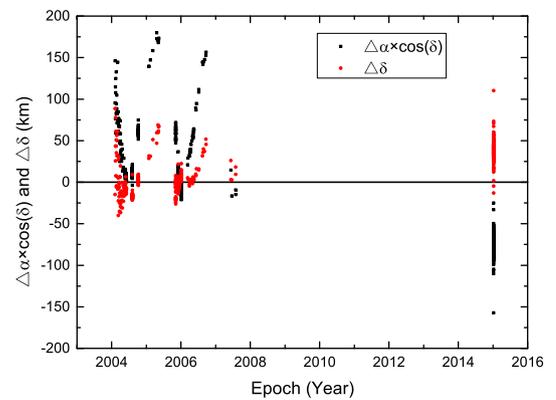

(c) Residuals relative to PH20

**Fig. 4.** The (O–C)s of positions of Phoebe in right ascension and declination relative to three different ephemerides: SAT375, SAT427 and PH20. Unit in km.

remind us that the big solar phase angle is a possible factor that cause big residuals. We should carefully correct the phase effect of

Phoebe. In our phase correction model, the shape of Phoebe is considered as a sphere with a radius of 106.4 km (Thomas et al.

**Table 4**
Mean values (mean) and standard deviations (std) of residuals of all 834 observed positions relative to the three different ephemerides.

|  | Sat375 | | Sat427 | | PH20 | |
| --- | --- | --- | --- | --- | --- | --- |
|  | mean | std | mean | std | mean | std |
| S(px) | 0.03 | 0.16 | 0.23 | 0.23 | 0.08 | 0.32 |
| L(px) | −0.00 | 0.13 | 0.11 | 0.56 | 0.17 | 0.44 |
| RA(″) | 0.08 | 0.19 | 0.87 | 0.65 | −0.19 | 0.90 |
| Dec(″) | −0.05 | 0.20 | −0.38 | 0.26 | 0.19 | 0.41 |
| RA(km) | 5.3 | 11.3 | 48.0 | 39.5 | −10.3 | 58.6 |
| Dec(km) | −2.2 | 10.9 | −20.7 | 17.1 | 13.9 | 23.7 |

(1) S (px): the residuals in column direction, unit in pixels.
(2) L (px): the residuals in Line direction, unit in pixels.
(3) RA(″): $\Delta\alpha \times \cos(\delta)$, unit in arc seconds.
(4) Dec(″): $\Delta\delta$, unit in arc seconds.
(5) RA(km): $\Delta\alpha \times \cos(\delta)$, unit in km.
(6) Dec(km): $\Delta\delta$, unit in km.

**Table 5**
The mean value and standard deviation of residuals in column and line of two classes of Phoebe positions relative to JPL ephemeris Sat375, unit in pixels.

|  | Non (CL1, CL2) | | (CL1, CL2) | |
| --- | --- | --- | --- | --- |
|  | Column | Line | Column | Line |
| Mean | 0.06 | 0.00 | −0.02 | −0.00 |
| Std | 0.18 | 0.15 | 0.13 | 0.09 |

(2018); Denk et al. (2018)). The method of phase correction given in Cooper et al. (2006) is applied. We know the sphere model is only an approximation of real Phoebe. In addition, it assumed that Phoebe's surface has a uniform brightness. In fact, Phoebe's surface has a significant variation of albedo. Hence, our phase correction model only partly removed the solar phase effect. Considering that phase errors can reach as much as 3 pixels, a more accurate model of Phoebe will obviously improve the centring of Phoebe. Because Table 3 gives the positions of Phoebe without phase correction, it is easy for users to replace our phase correction model with their own accurate one to get better results.

(3) As we all know, the chromatic aberration will bring positional measurement error of object. According to Liu et al. (2009), we estimate the positional error is small in NAC. But the real situation should be evaluated by experiment. It is outside the scope of the paper. In our results, we listed the situation of filters for each image. If the user has a model for fixing chromatic aberration, it will be easy to correct the position of Phoebe. On the other hand, the user can also select proper data to use instead of using all data.

In *Cassini* ISS observation, all these points above can produce errors of positional measurement. Especially the first two points have more influence on the accuracy of positional measurement than that in the earth-based observations, due to the close distance and no atmosphere.

## 5. Conclusions

Complementary astrometry of *Cassini* ISS images of Phoebe has been performed. All un-reduced ISS images of Phoebe between 2004 and 2017 have been considered. Finally, 834 ISS images of Phoebe have been reduced successfully by using *Caviar*. These images were taken between 2004 and 2015. During the astrometry, the modified moment method was used to obtain the centres of image stars and Phoebe. The Gaia EDR3 catalogue was used to get reference stars' positions. Those operations improved the measurement precision.

The final results show that the measurements fit well with JPL ephemeris SAT375. (O–C)s relative to SAT375 have means of 0.03 pixels in column and 0 pixels in line, with standard deviations of 0.16 and 0.13

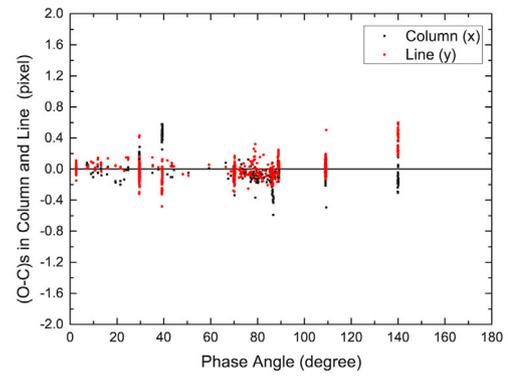

(a) Residuals in column and Line relative to SAT375 over solar phase angle

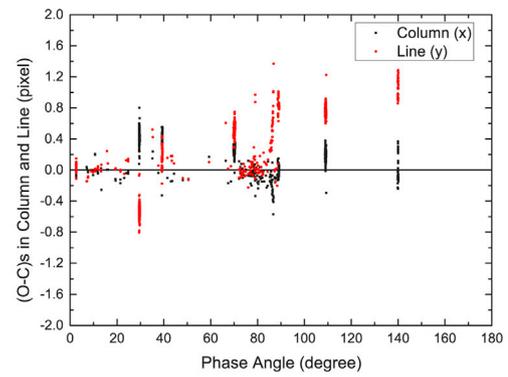

(b) Residuals in column and Line relative to SAT427 over solar phase angle

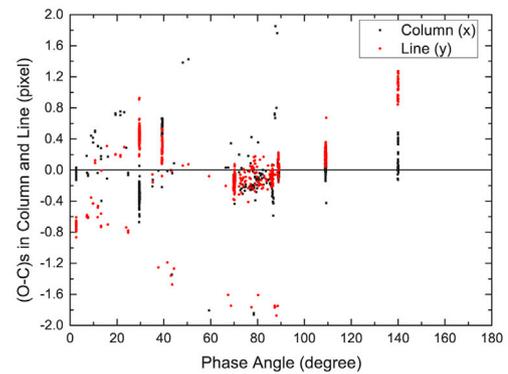

(c) Residuals in column and Line relative to PH20 over solar phase angle

**Fig. 5.** The distribution of Phoebe's positional residuals relative to three different ephemerides in column and line over solar phase angle.

pixels in column and line, respectively. In right ascension and declination, the means of these (O–C)s are 0.08″ and −0.05″, respectively. Their standard deviations are approximate 0.2″. In terms of residual in linear units, the means are about 5.3 km and −2.1 km in $\alpha$ and $\delta$, respectively. Their standard deviations are 11.3 km and 10.9 km, respectively.

Compared with the JPL ephemeris SAT427 and IMCCE ephemeris PH20, our measurements have a strong bias and a big dispersion. That suggests the earlier ephemeris SAT375 is the best fitting of our measurements among the three ephemerides.

**Data availability**

The full version of Table 3 can be found at ftp: 130.79.128.5 or https://cdsarc.unistra.fr/viz-bin/cat/J/other/P+SS, hosted at Strasbourg astronomical Data Center (Ochsenbein et al., 2000).

**CRediT authorship contribution statement**

**Q.F. Zhang:** Conceptualization, Software, Writing – original draft. **W.H. Qin:** Investigation, Writing – original draft. **Y.L. Ma:** Investigation, Writing – original draft. **V. Lainey:** Methodology, Writing – review & editing. **N.J. Cooper:** Methodology, Writing – review & editing. **N. Rambaux:** Methodology, Writing – review & editing. **Y. Li:** Investigation. **W.H. Zhu:** Investigation.

**Declaration of competing interest**

The authors declare that they have no known competing financial interests or personal relationships that could have appeared to influence the work reported in this paper.

**Acknowledgements**

This work has been partly supported by the Joint Research Fund in Astronomy under cooperative agreement between the National Natural Science Foundation of China and Chinese Academy of Sciences (No. U2031104), and National Natural Science Foundation of China (No. 11873026). This work has also been supported by the European Community's Seventh Framework Program (FP7/2007–2013) under grant agreement 263466 for the FP7-ESPaCE project. This work has made use of data from the European Space Agency (ESA) mission *Gaia* (https://www.cosmos.esa.int/gaia), processed by the *Gaia* Data Processing and Analysis Consortium (DPAC, https://www.cosmos.esa.int/web/gaia/dpac/consortium). Funding for the DPAC has been provided by national institutions, in particular the institutions participating in the *Gaia* Multilateral Agreement.

**Appendix A. Supplementary data**

Supplementary data to this article can be found online at https://doi.org/10.1016/j.pss.2022.105553.

**References**

Acton, C., Bachman, N., Semenov, B., Wright, E., 2018. A Look towards the Future in the Handling of Space Science Mission Geometry, vol. 150, pp. 9–12. https://doi.org/10.1016/j.pss.2017.02.013.

Acton, C.H., 1996. Ancillary Data Services of NASA's Navigation and Ancillary Information Facility, vol. 44, pp. 65–70. https://doi.org/10.1016/0032-0633(95)00107-7.

Arlot, J.E., Bec-Borsenberger, A., Fienga, A., Baron, N., 2003. Improvement of the ephemerides of Phoebe, 9th satellite of Saturn, from new observations made from 1995 to 2000. A&A 411, 309–312. https://doi.org/10.1051/0004-6361:20030985.

Castillo-Rogez, J.C., Johnson, T.V., Thomas, P.C., Choukroun, M., Matson, D.L., Lunine, J.I., 2012. Geophysical evolution of Saturn's satellite Phoebe, a large planetesimal in the outer Solar System. Icarus 219, 86–109. https://doi.org/10.1016/j.icarus.2012.02.002.

Cooper, N.J., Lainey, V., Meunier, L.E., et al., 2018. The Caviar software package for the astrometric reduction of Cassini ISS images: description and examples. A&A 610. https://doi.org/10.1051/0004-6361/201731713. A2.

Cooper, N.J., Murray, C.D., Lainey, V., Tajeddine, R., Evans, M.W., Williams, G.A., 2014. Cassini ISS mutual event astrometry of the mid-sized Saturnian satellites 2005-2012. A&A 572, A43. https://doi.org/10.1051/0004-6361/201424555 arXiv:1407.2045.

Cooper, N.J., Murray, C.D., Porco, C.C., Spitale, J.N., 2006. Cassini ISS astrometric observations of the inner jovian satellites, Amalthea and Thebe. Icarus 181, 223–234. https://doi.org/10.1016/j.icarus.2005.11.007.

Denk, T., Mottola, S., Tosi, F., Bottke, W.F., Hamilton, D.P., 2018. The irregular satellites of Saturn. In: Schenk, P.M., Clark, R.N., Howett, C.J.A., Verbiscer, A.J., Waite, J.H. (Eds.), Enceladus and the Icy Moons of Saturn, p. 409. https://doi.org/10.2458/azu_uapress_9780816537075-ch020.

Desmars, J., Li, S.N., Tajeddine, R., Peng, Q.Y., Tang, Z.H., 2013. Phoebe's orbit from ground-based and space-based observations. A&A 553, A36. https://doi.org/10.1051/0004-6361/201321114 arXiv:1303.0212.

Emelyanov, N., 2010. Precision of the ephemerides of outer planetary satellites. Planet. Space Sci. 58, 411–420. https://doi.org/10.1016/j.pss.2009.11.003.

Fraser, W.C., Brown, M.E., 2018. Phoebe: a surface dominated by water. AJNR 156, 23. https://doi.org/10.3847/1538-3881/aac213 arXiv:1803.04979.

Gaia Collaboration, 2016. The Gaia mission. A&A 595, A1. https://doi.org/10.1051/0004-6361/201629272 arXiv:1609.04153.

Gaia Collaboration, 2021. Gaia early data Release 3. Summary of the contents and survey properties. A&A 649, A1. https://doi.org/10.1051/0004-6361/202039657 arXiv:2012.01533.

Gomes-Júnior, A.R., Assafin, M., Braga-Ribas, F., Benedetti-Rossi, G., Morgado, B.E., Camargo, J.I.B., Vieira-Martins, R., Desmars, J., Sicardy, B., Barry, T., Campbell-White, J., Fernández-Lajús, E., Giles, D., Hanna, W., Hayamizu, T., Hirose, T., De Horta, A., Horvat, M., Hosoi, K., Jehin, E., Kerr, S., Machado, D.I., Mammana, L.A., Maybour, D., Owada, M., Rahvar, S., Snodgrass, C., 2020. The first observed stellar occultations by the irregular satellite Phoebe (Saturn IX) and improved rotational period. MNRAS 492, 770–781. https://doi.org/10.1093/mnras/stz3463 arXiv:1910.12188.

Gomes-Júnior, A.R., Assafin, M., Vieira-Martins, R., Arlot, J.E., Camargo, J.I.B., Braga-Ribas, F., da Silva Neto, D.N., Andrei, A.H., Dias-Oliveira, A., Morgado, B.E., Benedetti-Rossi, G., Duchemin, Y., Desmars, J., Lainey, V., Thuillot, W., 2015. Astrometric Positions for 18 Irregular Satellites of Giant Planets from 23 Years of Observations, vol. 580, p. A76. https://doi.org/10.1051/0004-6361/201526273 arXiv:1506.00045.

Jacobson, R.A., 2004. The orbits of the major saturnian satellites and the gravity field of Saturn from spacecraft and earth-based observations. AJNR 128, 492–501. https://doi.org/10.1086/421738.

Lainey, V., Casajus, L.G., Fuller, J., et al., 2020. Resonance locking in giant planets indicated by the rapid orbital expansion of Titan. Nat. Astron. 4, 1053–1058. https://doi.org/10.1038/s41550-020-1120-5 arXiv:2006.06854.

Lindegren, L., 1977. Meridian Observations of Planets with a Photoelectric Multislit Micrometer, vol. 57, pp. 55–72.

Liu, H.b., Tan, J.c., Huang, S.h., Li, X.j., Liu, L., Tan, W., 2009. Method for decrease of the centroid error of star image caused by stellar spectrum. In: Ye, S., Zhang, G., Ni, J. (Eds.), 2009 International Conference on Optical Instruments and Technology: Optoelectronic Measurement Technology and Systems, p. 75110U. https://doi.org/10.1117/12.839991.

Ochsenbein, F., Bauer, P., Marcout, J., 2000. The VizieR database of astronomical catalogues. AAS (Agents Actions Suppl.) 143, 23–32. https://doi.org/10.1051/aas:2000169 arXiv:astro-ph/0002122.

Owen Jr., W., 2003. Cassini ISS Geometric Calibration of April 2003. JPL IOM 312.E-2003.

Peng, Q., Vienne, A., Han, Y.B., Li, Z.L., 2004. Precise calibration of CCD images with a small field of view. Application to observations of Phoebe. A&A 424, 339–344. https://doi.org/10.1051/0004-6361:20034083.

Peng, Q.Y., Vienne, A., Zhang, Q.F., Desmars, J., Yang, C.Y., He, H.F., 2012. A convenient solution to geometric distortion and its application to Phoebe's observations. AJNR 144, 170. https://doi.org/10.1088/0004-6256/144/6/170.

Peng, Q.Y., Wang, N., Vienne, A., Zhang, Q.F., Li, Z., Meng, X.H., 2015. Precise CCD positions of Phoebe in 2011-2014. MNRAS 449, 2638–2642. https://doi.org/10.1093/mnras/stv469.

Peng, Q.Y., Zhang, Q.F., 2006. Precise positions of Phoebe determined with CCD image-overlapping calibration. MNRAS 366, 208–212. https://doi.org/10.1111/j.1365-2966.2005.09853.x.

Porco, C.C., Baker, E., Barbara, J., et al., 2005. Cassini imaging science: initial results on Phoebe and iapetus. Science 307, 1237–1242. https://doi.org/10.1126/science.1107981.

Porco, C.C., West, R.A., Squyres, S., et al., 2004. Cassini imaging science: instrument characteristics and anticipated scientific investigations at Saturn. Space Sci. Rev. 115, 363–497. https://doi.org/10.1007/s11214-004-1456-7.

Qiao, R.C., Tang, Z.H., Shen, K.X., et al., 2006. CCD astrometric observations of Phoebe in 2003-2004. A&A 454, 379–383. https://doi.org/10.1051/0004-6361:20054731.

Qiao, R.C., Xi, X.J., Dourneau, G., et al., 2011. CCD astrometric observations of Phoebe in 2005-2008. MNRAS 413, 1079–1082. https://doi.org/10.1111/j.1365-2966.2011.18214.x.

Rambaux, N., Castillo-Rogez, J.C., 2020. Phoebe's differentiated interior from refined shape analysis. A&A 643, L10. https://doi.org/10.1051/0004-6361/202039189.

Shen, K.X., Harper, D., Qiao, R.C., Dourneau, G., Liu, J.R., 2005. Re-determination of Phoebe's orbit. A&A. 437, 1109–1113. https://doi.org/10.1051/0004-6361:20052728.

Shen, K.X., Li, S.N., Qiao, R.C., et al., 2011. Updated Phoebe's orbit. MNRAS 417, 2387–2391. https://doi.org/10.1111/j.1365-2966.2011.19420.x.

Simonelli, D.P., Kay, J., Adinolfi, D., et al., 1999. Phoebe: albedo map and photometric properties. Icarus 138, 249–258. https://doi.org/10.1006/icar.1999.6077.

Tajeddine, R., Cooper, N.J., Lainey, V., Charnoz, S., Murray, C.D., 2013. Astrometric reduction of Cassini ISS images of the saturnian satellites mimas and enceladus. A&A 551, A129. https://doi.org/10.1051/0004-6361/201220831.

Tajeddine, R., Lainey, V., Cooper, N.J., Murray, C.D., 2015. Cassini ISS astrometry of the saturnian satellites: tethys, dione, rhea, iapetus, and Phoebe 2004-2012. A&A 575, A73. https://doi.org/10.1051/0004-6361/201425605.

Thomas, P.C., Tiscareno, M.S., Helfenstein, P., 2018. The inner small satellites of Saturn, and hyperion. In: Schenk, P.M., Clark, R.N., Howett, C.J.A., Verbiscer, A.J., Waite, J.H. (Eds.), Enceladus and the Icy Moons of Saturn, p. 387. https://doi.org/10.2458/azu_uapress_9780816537075-ch019.

Veiga, C.H., Vieira Martins, R., Andrei, A.H., 2000. CCD observations of Phoebe. AAS (Agents Actions Suppl.) 142, 81–84. https://doi.org/10.1051/aas:2000138.

Zhang, Q.F., Lainey, V., Cooper, N.J., Vienne, A., Peng, Q.Y., Xiong, Y.T., 2018a. First astrometric reduction of Cassini Imaging Science Subsystem images using an automatic procedure: application to Enceladus images 2013-2017. MNRAS 481, 98–104. https://doi.org/10.1093/mnras/sty2187.

Zhang, Q.F., Lainey, V., Vienne, A., Cooper, N.J., Peng, Q.Y., Wang, N., 2018b. Astrometric reduction of Cassini ISS images of enceladus in 2015 based on Gaia DR1. In: Recio-Blanco, A., de Laverny, P., Brown, A.G.A., Prusti, T. (Eds.), Astrometry and Astrophysics in the Gaia Sky, pp. 411–412. https://doi.org/10.1017/S1743921317005555.

Zhang, Q.F., Zhou, X.M., Tan, Y., et al., 2021. A comparison of centring algorithms in the astrometry of Cassini imaging science subsystem images and Anthe's astrometric reduction. MNRAS 505, 5253–5259. https://doi.org/10.1093/mnras/stab1626.